\documentclass{jfm}
\usepackage{graphicx}
\usepackage{epstopdf, epsfig}
\usepackage{caption}
\usepackage{subcaption}
\usepackage[export]{adjustbox} 
 \usepackage{floatrow}
\usepackage{siunitx}
\usepackage{amsmath}
\usepackage{cleveref}

\shorttitle{Simulation of Electrospray Emission Processes for Highly Conductive Liquids}
\shortauthor{Henry Huh and Richard E. Wirz}

\title{Simulation of Electrospray Emission Processes for Highly Conductive Liquids}

\author{Henry Huh\aff{1}
 \and Richard E. Wirz\aff{1}}

\affiliation{\aff{1}Department of Mechanical and Aerospace Engineering, University of California, Los Angeles, CA 90095, USA}

\begin{document}

\maketitle

\begin{abstract}

An electrohydrodynamic numerical model is used to explore the electrospray emission behavior of both moderate and high electrical conductivity liquids under electrospray conditions. The Volume-of-Fluid method, incorporating a leaky-dielectric model with a charge relaxation consideration, is used to conserve charge to accurately model cone-jet formation and droplet breakup. The model is validated against experiments and agrees well with both droplet diameters and charge-to-mass ratio of emitted progeny droplets. The model examines operating conditions such as flow rate and voltage, with fluid properties also considered, such as surface tension, electrical conductivity, and viscosity for both moderate and high conductivity. For high conductivity and surface tension, the results show that high charge concentration along with the meniscus and convex cone shape results in a higher charge-to-mass ratio of the emitted droplets while lower conductivity and surface tension tend towards concave cone shapes and lower charge-to-mass droplets. Recirculation flows inside the bulk liquid are investigated across a range of non-dimensional flow rates, $\delta$, and electric Reynolds numbers, $Re_E$. For high conductivity liquid emission at the minimum stable flow rate, additional recirculation cells develop near the cone tip suggesting the onset of the axisymmetric instability.
\end{abstract}


\section{Introduction}\label{sec1}

Electrospray is an attractive technique for many applications, such as mass spectrometry, MEMS fabrication, nano-fiber deposition, and electric propulsion (EP). In the last few decades, considerable study has been undertaken to understand the underlying physics of electrospray electrohydrodynamics(EHD). For electrospray operation, an electric potential difference of hundreds to several thousand volts is applied between an emitter and an extraction electrode, producing a cone-shaped meniscus that leads to a liquid jet, droplets, and ions with relatively high velocity due to the E-field concentration overall large potential drop. Changing operating conditions like flow rate, voltage, and physical properties of the liquid such as electrical conductivity, permittivity, surface tension, and viscosity allows for several different emission modes. Pioneering work was done from \cite{Zeleny_first} by observing numerous modes of the cone-jet. Among the modes, steady cone-jet mode is of high interest for its stable emission of the liquid. \cite{taylor} explained the steady conical meniscus by balancing surface tension and electrostatic forces. The Ohmic, ``leaky-dielectric'' model suggested by \cite{melcher} describes the role of tangential electrostatic forces acting on a liquid interface in generating a cone shaped meniscus and progeny droplets. This theoretical model deviates from the perfect dielectric, conductive model where the electrostatic force acts only normal to the surface. The leaky dielectric model is a plausible approach in that although the bulk liquid is free of charge, charge accumulation at the liquid surface allows tangential stress to deform the fluid. Saville then introduced electrokinetic physics, complementary to the work of the Taylor-Melcher leaky-dielectric model (\cite{saville}). Saville's complex, non-linear physics led to a description of unsteady cone-jet behavior, and the formation of both main and satellite droplets, which are difficult to observe. In an effort to elucidate the complex physics of electrospray emission, important scaling laws have been developed under various assumptions (\cite{ganan_montanero2009,ganan_scaling,GANANCALVOscaling1997,delamora}). Although the initial conditions and the resultant droplet measurements can be defined with scaling laws, the emission mechanism during the evolution of a cone-jet is not described. Such insight requires high-fidelity EHD simulations and by using this approach we can further investigate the physics of electrospray emission evolution, and compare results with scaling laws and experimental observations.

Electrospray emission behavior can be characterized by the various forces acting at the meniscus of the cone. In general, a normal electrostatic force balances a pressure jump across the interface. A tangential electrostatic force along the meniscus leads to cone formation, and eventually jetting and droplet breakup if the electrostatic force surpasses the finite surface tension force. With a more polarizable liquid, the polarization force behaves as a driving force to accelerate the jet, eventually causing polarization-driven instability (\cite{melcher_1981, ganan_polar}). 

In the process of emission, the cone-to-jet region can be defined by the dominant contribution to the total emitting current being from Ohmic conduction to surface charge convection. Note that the "charge relaxation region", defined by \cite{delamora}, is the same as the cone-to-jet region, in that the hydrodynamic and electrical relaxation time scales are of the same order of magnitude in these regions. Electrohydrodynamic flow behavior is commonly characterized by the comparison between charge relaxation time $t_e=\varepsilon/\sigma$, and hydrodynamic characteristic time, $t_h=LR^2/Q$, where $\varepsilon$ is permittivity, $\sigma$ is conductivity, $L$ is characteristic length, $R$ is emitter diameter, and $Q$ is flow rate. Later in this paper we show that this is only true for low conductivity liquids. 

 Ever since \cite{Zeleny_instability} observed hydrostatic flow instabilities, fluid behaviors and the structure of the cone-jet have also been of keen interest and has been analyzed predominantly for low conductivity liquids. In particular, recirculation flow was first observed experimentally by \cite{hayati} and continuously researched both experimentally and numerically by a variety of research groups. \cite{shtern_striking} observed swirl motion due to flow instability and claimed axisymmetric breakdown as a reason for the swirl motion. \cite{herrada,dastourani,ganan_recirculation} have used numerical solutions to estimate the behavior of recirculation when increasing the flow rate. \cite{cherney_1999_structure} derived asymptotic relations to analyze the shape, key distributions, electric field, and surface charge of the cone-jet meniscus. Recirculation flow inside a bulk liquid is mainly due to a driving tangential electrostatic stress at the fluid surface. The onset of recirculation phenomena has been observed both experimentally and numerically with decreasing flow rate near the minimum stable flow rate\cite{ganan_rec,ganan_recirculation}. 

Several numerical models of electrospray emission have been developed over the years to describe the cone-jet. The boundary element/integral method (BEM) used in several electrospray studies is computationally cost-efficient and performs accurate analysis. However, it is heavily constrained to linear problems and is reduced to one dimension at the liquid interface, which limits the observation of flow instability inside the bulk liquid (e.g., recirculation flow) or calculation of droplet specific charge (which is an important observable parameter for electric propulsion). The finite volume method (FVM), on the other hand, is powerful for robust handling of nonlinear conservation equations that appear in transport problems. Several EHD models have been developed based on FVM. For example, \cite{lopez, herrada} used Volume-of-Fluid analysis to track interfaces within the framework of an open-source tool, Gerris specifically developing multiphase problems (\cite{popinet}). \cite{ivo} then developed an EHD code based on the work of \cite{lopez} using the OpenFOAM framework and was extended by \cite{dastourani} to generate physical insights into electrospray emission physics for low conductivity liquid. Moderate-to-high conductivity modeling, on the other hand, requires additional numerical considerations and treatments to accurately resolve jet breakup and droplet formation. The objective of this paper is to present and EHD model developed specifically for simulating electrospray emission of moderate to high conductivity liquids. Our overarching approach develops EHD code in OpenFOAM with several modifications to the EHD governing equations and the model from \cite{ivo}, in order to capture both the cone-formation and droplet breakup physics while conserving charge for moderate and high conductivity liquids.


In this paper, we first validate the suggested model where several important modifications are made to accommodate higher conductivity liquids. For example, by considering the charge relaxation for higher conductivity liquids, the charge transport equation is modified to effectively conserve charge in the cone meniscus and the emitted droplets. Then we investigate the charge distribution along the meniscus over a range of operating conditions. In particular, we run the model across a range of fluid properties that are critical in defining the steady cone-jet mode to show how the meniscus shape is determined by the various conditions showing that charge concentration along the meniscus subsequently affects the cone-to-jet length and the charge-to-mass ratio of emitted droplets. We then study the recirculation phenomenon for both moderate and high conductivity liquids and find that the high conductivity liquid exhibits additional recirculation cells at the cone-tip near the minimum stability flow rate. We define a cone-to-jet region over a length $L_{cj}$, to be the region over which the charge density changes from 5 to 95 \% of its final value. The charge density distribution along this cone-to-jet region, and the resulting charge-to-mass ratio of emitted droplets, is observed for various flow conditions (flow rate, voltage) and fluid properties (electrical conductivity, surface tension, viscosity) that are critical to emission flow behavior. 
The results from these parametric evaluations reveal intriguing cone and jet flow phenomena that lead to experimentally observed trends in important electrospray metrics such as droplet size and charge-to-mass ratio. 

The paper is organized as follows: First, we discuss the governing equations in Section~\ref{sec2}. The modeling results for moderate conductivity liquid is presented in Section~\ref{sec3.1} and high conductivity liquid in Section~\ref{sec3.2}. Concluding remarks are provided in Section~\ref{sec4}.








\section{Model Formulation}\label{sec2}


\subsection{Fluid flow}

The simulations of fluid flow for both low and high conductivity liquid ranges are governed by the incompressible EHD continuity and momentum equations  
\begin{equation}\label{eq:2.1}
  \bnabla\bcdot\boldsymbol{u} = 0,
\end{equation}
\begin{equation}\label{eq:2.2}
  \rho \Big[\frac{\partial\boldsymbol{u}}{\partial t}+\boldsymbol{u}\bcdot\bnabla\boldsymbol{u}\Big] = -\bnabla\boldsymbol{P} + \mu \bnabla^2 \boldsymbol{u} + \rho\boldsymbol{g} + \boldsymbol{F}_E + \boldsymbol{F}_{ST},
\end{equation}
where the electrostatic force $\boldsymbol{F}_{E}$, and surface tension force $\boldsymbol{F}_{ST}$ are added to the Navier-Stokes equation for density $\rho$, pressure $P$, dynamic viscosity $\mu$, and velocity $\boldsymbol{u}$. The local surface tension force $\boldsymbol{F}_{ST}$ is determined to be a volumetric force by the continuum surface force (CSF) model developed by \cite{BRACKBILL1992335}.:
\begin{equation}\label{eq:2.3}
  \boldsymbol{F}_{ST} = \gamma\kappa\boldsymbol{n}=\gamma(-\bnabla\bcdot\hat{\boldsymbol{n}})\boldsymbol{n},
\end{equation}
for surface tension coefficient $\gamma$, interface curvature $\kappa$, and unit normal vector $\hat{n}$.

The proposed model uses a Volume-Of-Fluid (VOF) method to capture the interface between liquid and vacuum by the transport equation in Eq.~\ref{eq:2.4}:

\begin{equation}\label{eq:2.4}
\frac{\partial\alpha}{\partial t} + \bnabla\bcdot(\boldsymbol{u}\alpha) = 0.
\end{equation}
where $\alpha$  is volume fraction, and $t$ is time.

\subsection{Electrostatics}
The model is governed by electrohydrodynamics as magnetic phenomena are negligible under the given conditions. Maxwell’s equations are reduced to the electrostatic equation in Eq.~\ref{eq:2.5} and Gauss’s law for Eq.~\ref{eq:2.6}, where $\boldsymbol{E}$ is electric field, $\varepsilon$ is electrical permittivity and $\rho_e$ is volumetric charge density:
\begin{equation}\label{eq:2.5}
  \bnabla\times\boldsymbol{E}=0;
\end{equation}
\begin{equation}\label{eq:2.6}
  \bnabla\bcdot(\varepsilon\boldsymbol{E})=\rho_e.
\end{equation}

The charge conservation equation~\ref{eq:2.7} is derived as~\ref{eq:2.8} by substituting the current density with both the Ohmic charge conduction and charge convection transport, $\boldsymbol{J}=\sigma\boldsymbol{E}+\rho_e\boldsymbol{u}$.

\begin{equation}\label{eq:2.7}
 \frac{\partial\rho_e}{\partial t}+ \bnabla\bcdot\boldsymbol{J}=0,
\end{equation}

\begin{equation}\label{eq:2.8}
  \frac{\partial\rho_e}{\partial t}+\bnabla\bcdot(\rho_e\boldsymbol{u}) = -\bnabla\bcdot(\sigma\boldsymbol{E}),
\end{equation}

Gauss’s law~\ref{eq:2.6} and the electrostatic charge conservation equation~\ref{eq:2.8} calculate the charge density and electric field.
 
For low conductivity (\SI{\sim e-9}{\siemens \per \meter}), the charge relaxation time, $\tau_{e}=\dfrac{\epsilon}{\sigma}$ is \SI{\sim e-2}{\second}, which allows the speed of the fluid flow to be comparable to that of the charge motion within the fluid. The charge convection then has a non-negligible effect on the charge transport. This effect is well-observed in droplet deformation in the presence of an electric field for liquids of conductivity less than \SI{\sim e-9}{\siemens \per \meter}, where charge convection tends to enhance prolate deformation due to charge concentration (\cite{sengupta}). The charge relaxation times observed in experiments (\cite{tang,salipante}) are less than \SI{2.5}{\second} and convective effects are observable below an electrical conductivity of \SI{\sim e-8}{\siemens \per \meter}, which is significant even at charge relaxation times of \SI{\sim e-3}{\second}~(\cite{feng}). Also convective effects on droplet deformation are all reported in the low conductivity range, where the conductivity ratio of the liquid and the medium is \SIrange{0}{100}~(\cite{lanauze,sengupta}).
 
For moderate (\SI{\sim e-7}{\siemens \per \meter}) and high (\SI{\sim e-4}{\siemens \per \meter}) electrical conductivity, the charge relaxation time ranges from \SI{\sim e-4}{\second} to \SI{\sim e-1}{\second}, and is significantly smaller than the hydrodynamic time scale (\cite{ganan_scaling}). This indicates that charge conduction dominates over charge convection for charge transport inside the fluid. Resultant jet velocities are approximately \SI{100}{\meter \per \second} in experiments (\cite{gamero}) at the given conductivity where numerical results without charge convection share similar results indicating that charge conduction is the major contributor to jet formation due to fast relaxation. It is therefore concluded that charge convection has a negligible effect on charge transport for moderate-to-high conductivity electrosprays. We will validate our analysis with other key outputs throughout the paper. 

Based on the discussion above, and because the model developed herein focuses on moderate and high conductivity liquids, the charge convection term in Eq~\ref{eq:2.8} can then be neglected, resulting in Eq~\ref{eq:2.9}:

\begin{equation} \label{eq:2.9}
  \frac{\partial\rho_e}{\partial t} + \bnabla\bcdot\sigma\boldsymbol{E}=0.
\end{equation}
Note that Eq.~\ref{eq:2.9} can also be derived from an electrokinetic analysis~(\cite{herrada}).

Electrostatic force is governed by the sum of Coulombic and polarization forces, per Eq.~\ref{eq:2.10}, and acts on the liquid surface.

\begin{equation}\label{eq:2.10}
  \boldsymbol{F}_E = \rho_e\boldsymbol{E} - \frac{1}{2}\boldsymbol{E}^2\bnabla\varepsilon.
\end{equation}


Higher conductivity electrosprays are typically run in near vacuum conditions wehre micro- and nano-sized droplets are generated in the emission region due to the forces along the cone and jet meniscus. In order to capture the interface of a the cone and jet, the Volume-of-Fluid method is employed for two-phase flow simulations.

Preliminary simulation results from our model showed that an important consideration for high conductivity liquids is to accurately treat the liquid-vacuum interface to avoid numerically-induced charge and mass leakage. Therefore, to treat and discretize the two fluids, the $\sigma$, $\varepsilon$, and $\rho$ value of each calculation cell is determined based on the volume fraction of the cell, $\alpha$. \cite{TOMAR20071267} used an arithmetic mean of the given properties: however, this lead to significant numerical diffusion in our simulations for higher conductivity liquids. Therefore, in this study, we have devised a new way of calculating the cell properties based on the liquid and vacuum properties, that results in better model validation. By implementing model properties of Eq.~\ref{eq:2.11} and \ref{eq:2.12}, steeper transition of the properties is applied to the governing equations~\ref{eq:2.6}, \ref{eq:2.9}, and \ref{eq:2.10}; which allows superior conservation of charge within the volume fraction compared to use an arithmetic mean.

\begin{equation}\label{eq:2.11}
 \sigma_{cell} = \Big(\alpha_{liq}\sigma_{liq}^{1/10} + \alpha_{vac}\sigma_{vac}^{1/10}\Big)^{10},
\end{equation}

\begin{equation}\label{eq:2.12}
 \varepsilon_{cell} = \Big(\alpha_{liq}\varepsilon_{liq}^{1/10} + \alpha_{vac}\varepsilon_{vac}^{1/10}\Big)^{10},
\end{equation}


Additionally, in order to reduce numerical diffusion, a 2nd order accurate linear upwind difference scheme was employed.
As described in Section 3, full-scale computational domains are used for both moderate conductivity (Section 3.1) and high conductivity (Section 3.2) liquid emission simulations based on published experimental results and setups described by~\cite{tang} and~\cite{gamero}, respectively. The properties for the fluids used in these references and in our simulations are described in Table 1. For the simulations in Section 3, a total pressure of \SI{0}{\pascal} is used at the wall to set a vacuum condition inside the chamber. Also, the velocity and electrical potential are set to a zero gradient boundary condition at the wall. 

\begin{figure}
  \centerline{\includegraphics[scale=0.35]{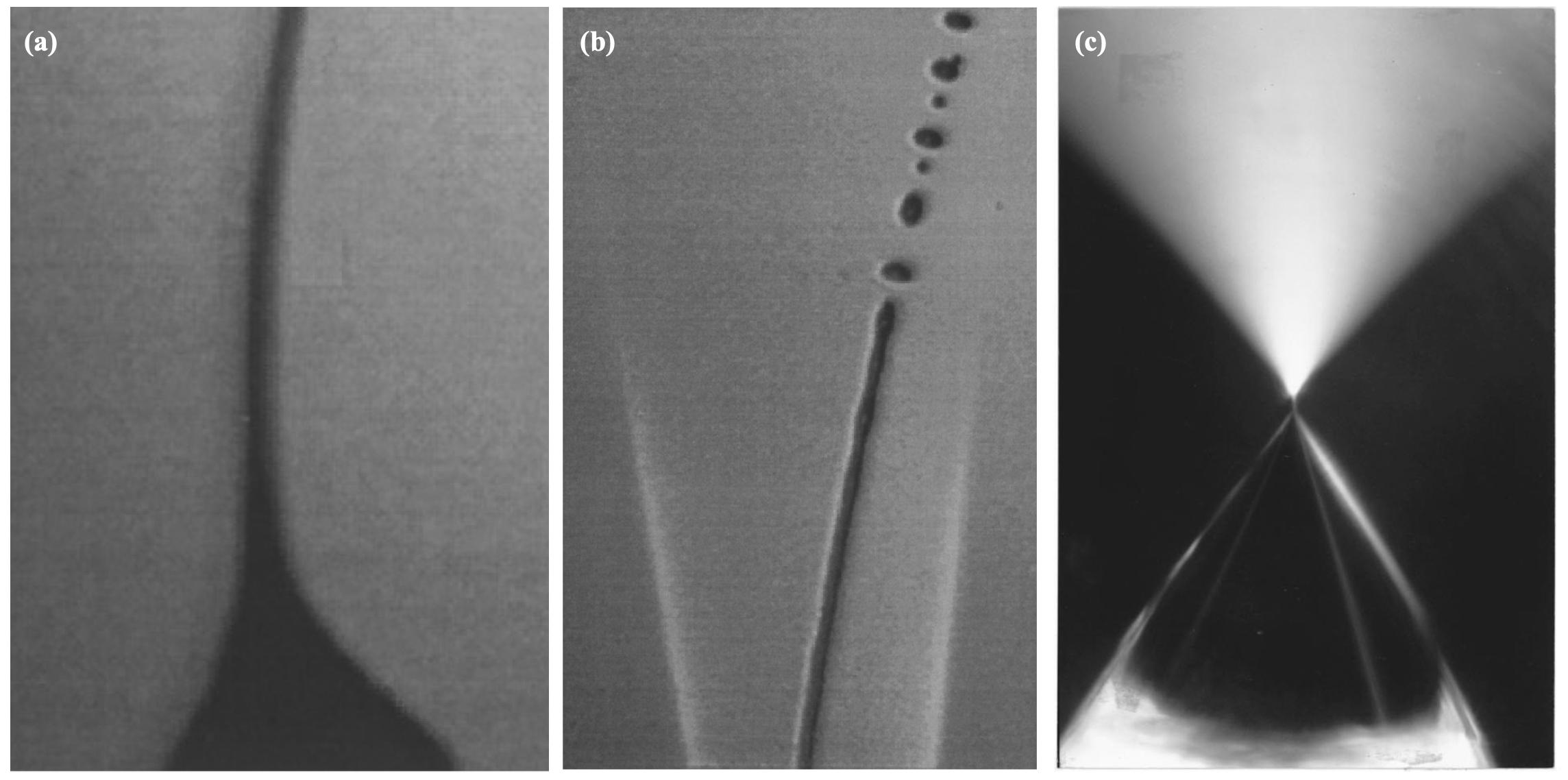}}
  \caption{Experimental electrospray emission results for (a) moderate conductivity, heptane $\sigma$=~\SI{6.26e-7}{\siemens \per \meter}, (b) subsequent jet and emitted droplets from (a), and (c) high conductivity, tributyl phosphate $\sigma$=~\SI{0.033}{\siemens \per \meter}} (\cite{tang,gamero_direct_measurement}).
\label{fig:kd1}
\end{figure}

Output parameters are investigated as a function of a non-dimensional flow rate in Eq.~\ref{eq:2.14}, and electric Reynolds number in Eq.~\ref{eq:2.15} where $Q$ is flow rate, $\varepsilon_0$ is permittivity of free space. Figure~\ref{fig:kd1} shows experimental observations of (a,b) moderate and (c) high conductivity liquids where emission behaviors are different. Here we present modeling results to reproduce the observations of Figure~\ref{fig:kd1} at the given operating conditions and geometrical configuration, that qualitatively and quantitatively validate the modeling results. 
 
 \begin{equation}\label{eq:2.14}
 \delta = \frac{\rho \sigma Q}{\gamma \varepsilon_0},
\end{equation}

\begin{equation}\label{eq:2.15}
    Re_E=(\frac{\rho \varepsilon_0 \gamma^2}{\mu^3 \sigma})^{1/3}.
\end{equation}

\section{Results and Discussion}

\begin{table}
  \begin{center}
\def~{\hphantom{0}}
  \begin{tabular}{lccccc}
      Liquid & $\rho {\normalfont(kg/m^3)}$ & $\sigma {\normalfont(S/m)}$ & $\gamma {\normalfont(N/m)}$ & $\varepsilon {\normalfont(F/m)}$ & $\mu {\normalfont(Pa S)}$ \\
       Heptane & 684 & $6.26\times10^{-7}$ & 0.0186 & 1.91 & $4.28\times10^{-4}$\\
       TBP & 976 & $2.3\times10^{-4}$ & 0.028 & 8.91 & $3.59\times10^{-3}$\\
    
  \end{tabular}
  \caption{Liquid properties of heptane and tributyl phosphate(TBP)}
  \label{tab:kd}
  \end{center}
  
\end{table}

\begin{figure}
  \centerline{\includegraphics[scale=0.35]{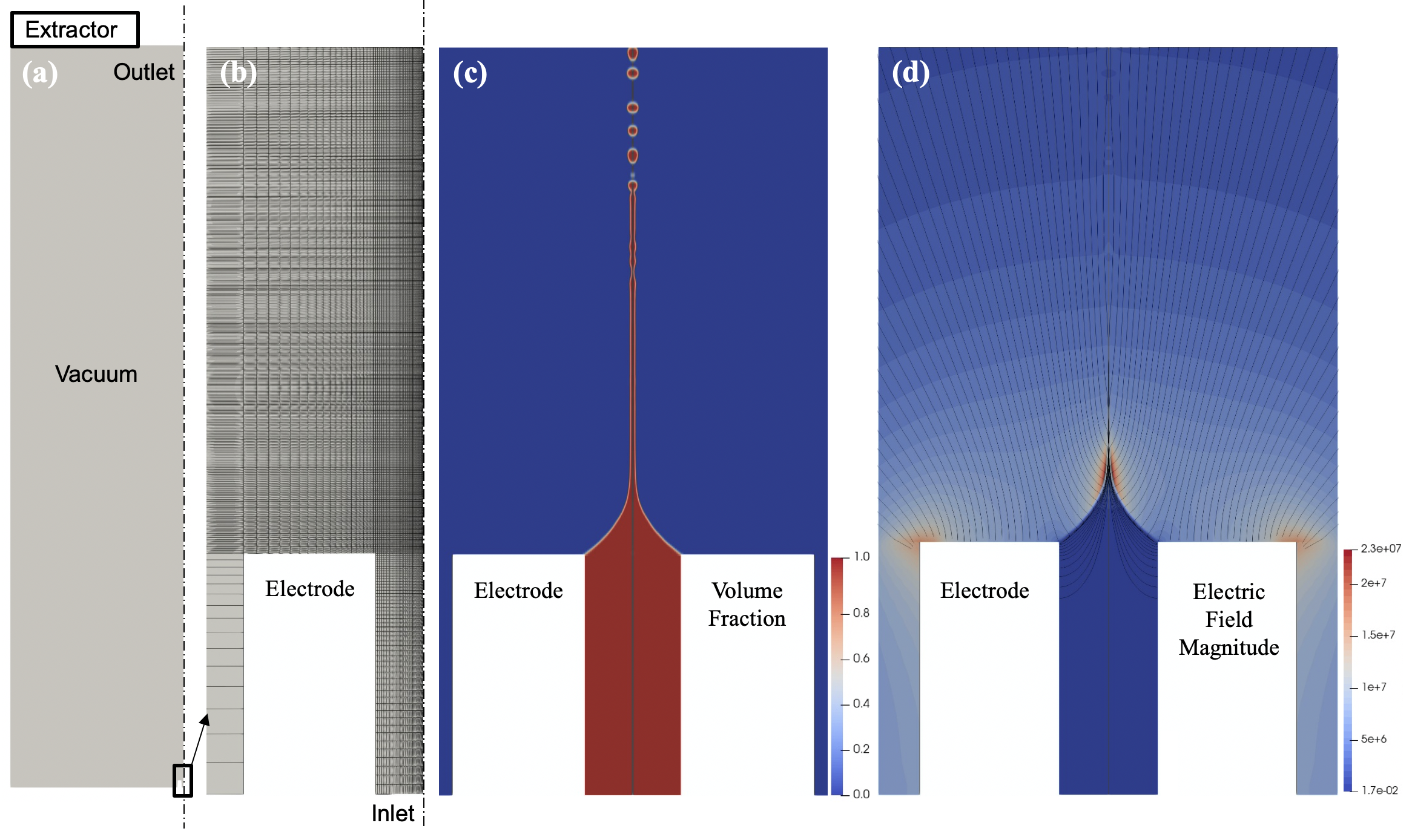}}
  \caption{Electrospray emission results for moderate conductivity ($\sigma=$~\SI{6.26e-6}{\siemens \per \meter}) (a) 2D axisymmetric full computational domain with 138,800 mesh cells (b) Magnified computational domain near emission region (c) Volume fraction result at steady state ($t=$~\SI{0.5}{\milli\second}) (d) Electric field magnitude}
\label{fig:kd2}
\end{figure}

\begin{figure}
  \centerline{\includegraphics[scale=0.25]{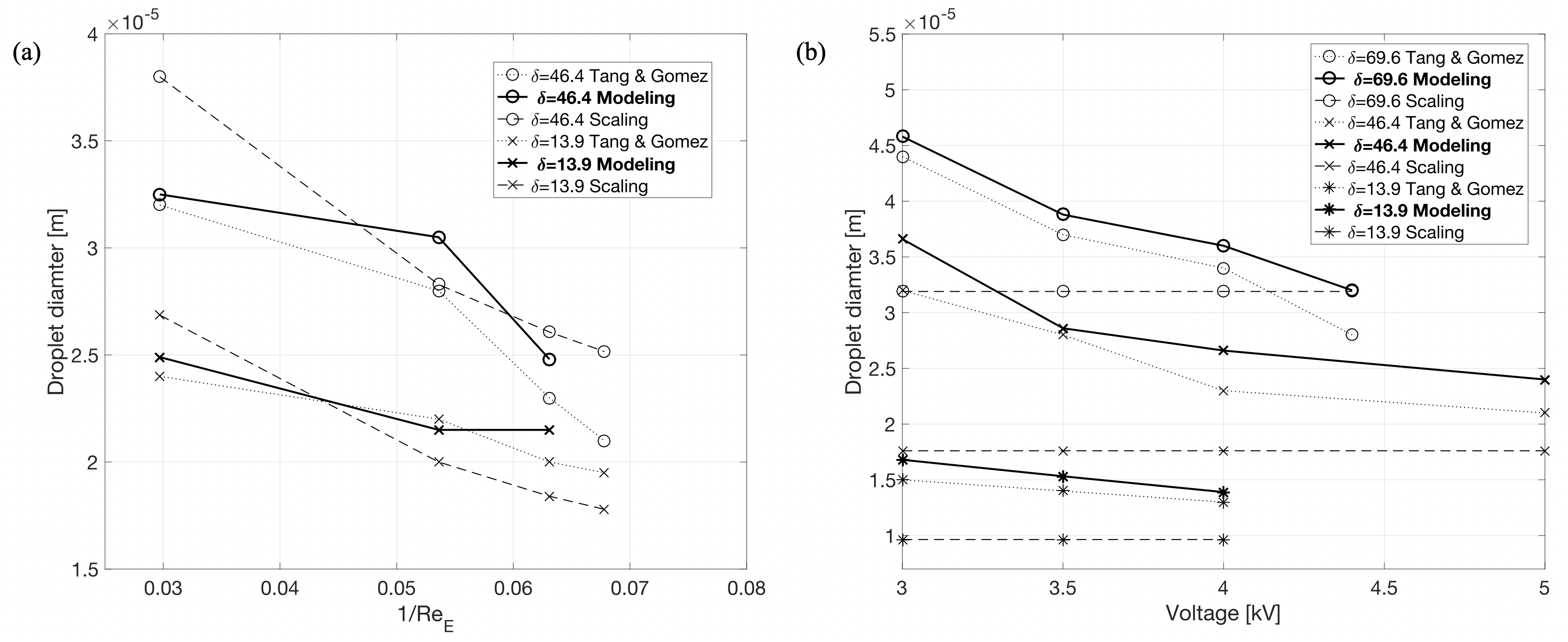}}
 \caption{Quantitative validation of the moderate electrical conductivity liquid Heptane. (\textit{Left}) Droplet diameter measurements with voltage ranging from \SIrange{3}{5}{\kilo\volt}. (\textit{Right}) Droplet diameter with different electrical conductivities}
 \label{fig:kd3}
\end{figure}

\subsection{Moderate conductivity}\label{sec3.1}

\begin{figure}
  \centerline{\includegraphics[scale=0.32]{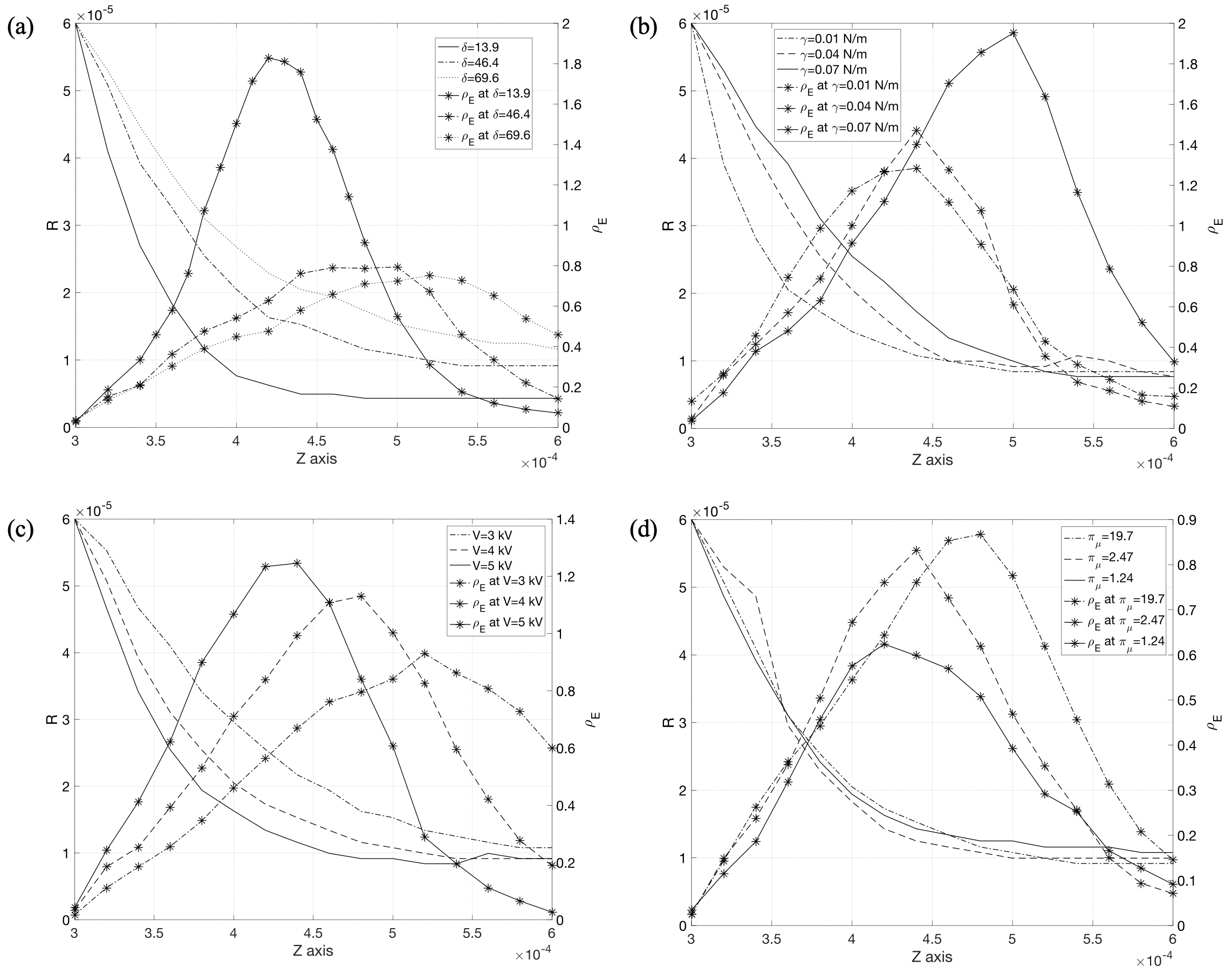}}
  \caption{
  (a) Meniscus and charge density distribution at different $\delta$, (b) surface tension coefficient, (c) voltage, (d) viscosity.}
\label{fig:kd4}
\end{figure}

\begin{figure}
  \centerline{\includegraphics[scale=0.28]{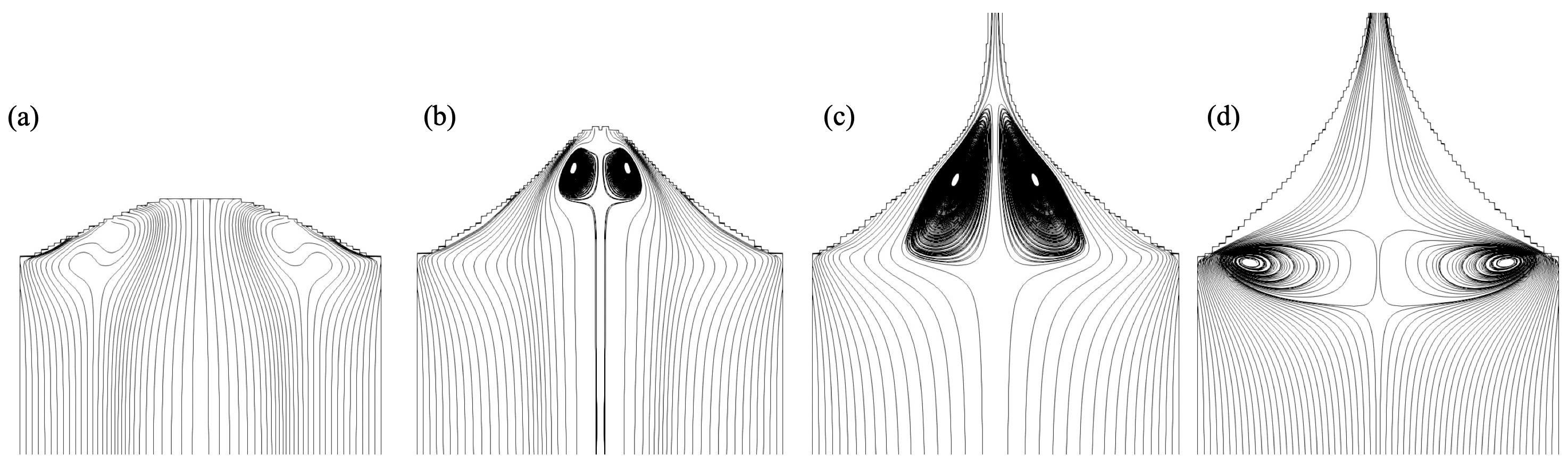}}
  \caption{Electrospray emission evolution with recirculation flow inside the bulk liquid, with moderate conductivity at $Q=$~\SI{5e-10}{\meter \cubed \per \second}, $V=$~\SI{4}{\kilo\volt}. Time~= (a) \SI{0.19}{\milli\second} (b) \SI{0.28}{\milli\second} (c) \SI{0.34}{\milli\second} (d) \SI{0.5}{\milli\second}.}
\label{fig:kd5}
\end{figure}

For the moderate conductivity case shown in Figure~\ref{fig:kd2} following the experimental setup from~\cite{tang}, the nozzle inner and outer diameters are \SI{120}{\micro \meter} and \SI{450}{\micro \meter} respectively, with the emitter-extractor length at \SI{29.8}{\milli \meter}, and extractor orifice diameter of \SI{12}{\milli\meter}. 
 
 In Figure~\ref{fig:kd2}, simulation results of volume fraction and electric field are presented for a moderate conductivity liquid, showing good qualitative agreement with the experimental result of Figure~\ref{fig:kd1}. It is observed in Figure~\ref{fig:kd2}(d) that the peak of the electric field magnitude appears at the cone-to-jet region due to high charge accumulation. Note that the maximum electric field magnitude at the cone-to-jet region is well below the minimum threshold for ion emission in Figure~\ref{fig:kd2}(d) for moderate conductivity and that the model does not account for ion emission(\cite{gamero_direct_measurement,garoz_ionemission,romero_ionemission}). In Figure~\ref{fig:kd3}, first-emitted droplets at steady state measured from the numerical results are compared with the universal scaling laws of~\cite{ganan_scaling}, and experimental results from~\cite{tang}. In Figure~\ref{fig:kd3}(a), $1/Re_E$ is varied, and in Figure~\ref{fig:kd3}(b) the flow rate is varied from \SIrange{5e-10}{2.5e-9}{\meter\cubed\per\second}, and voltage varied from \SIrange{3}{5}{\kilo\volt}. Good quantitative agreement is found between the model and experimental observations. The droplet size tends to decrease with decreasing $Re_E$, decreasing flow rate, and increasing voltage as already discussed in the literature~(\cite{ganan_scaling,GANANCALVOscaling1997,tang,dastourani,delamora,gamero}). The dependence of droplet size on voltage tends to decrease with lower flow rate and higher conductivity in Figure~\ref{fig:kd3}(b) verifying the observations from \cite{tang}. This is mainly due to a charge accumulation effect downstream at the cone-tip that dominates the electric potential at the electrode, and changes the emission behavior. In Figure~\ref{fig:kd4}, charge density distributions at various flow conditions and fluid parameters are shown along the meniscus where the charge distribution is highly dependent on the shape of the meniscus. Decreasing the flow rate allows for a steeper meniscus, followed by a higher maximum charge density at the cone-to-jet region, with shorter $L_{cj}$. Note that $\rho_E$ is a normalized surface charge density with the universal expression derived from \cite{ganan_scaling}. Increasing the emitter voltage lets the electric field magnitude increase at the emitter tip. $\delta=$~\SI{13.9} in Figure~\ref{fig:kd4}(a) has the highest charge accumulation and allows for a steeper meniscus with finer jet that leads to smaller droplets. Increasing the surface tension of the liquid also gives high charge density profiles and flattening of the meniscus as shown in Figure~\ref{fig:kd4}(b). The voltage increase in Figure~\ref{fig:kd4}(c) allows a higher electrostatic force along the meniscus that allows similar sensitivity trends to that of decreasing $\delta$. In Figure~\ref{fig:kd4}(d), the meniscus shape and the charge density are shown with various viscous dimensionless parameters adopted from \cite{GANANCALVOscaling1997}. As shown in the Figure, despite the fact that the viscosity change has a non-negligible effect on different modes of the emission, viscosity changes do not make much difference to the meniscus shape or charge density.
 




Emission evolution with time is shown in Figure~\ref{fig:kd5}, with progressive recirculation flow behavior observed at the cone-to-jet region due to the tangential Coulombic force along the conical meniscus. Note that the recirculation cell develops from the tip at $t=$~\SI{0.28}{\milli\second} then progresses upstream near the inner tip of the electrode at $t=$~\SI{0.5}{\milli\second}, and stays in steady cone-jet mode, which agrees with past experiments of moderate conductivity liquids (\cite{ganan_rec}). 



\subsection{High Conductivity}\label{sec3.2}

\begin{figure}
  \centerline{\includegraphics[scale=0.35]{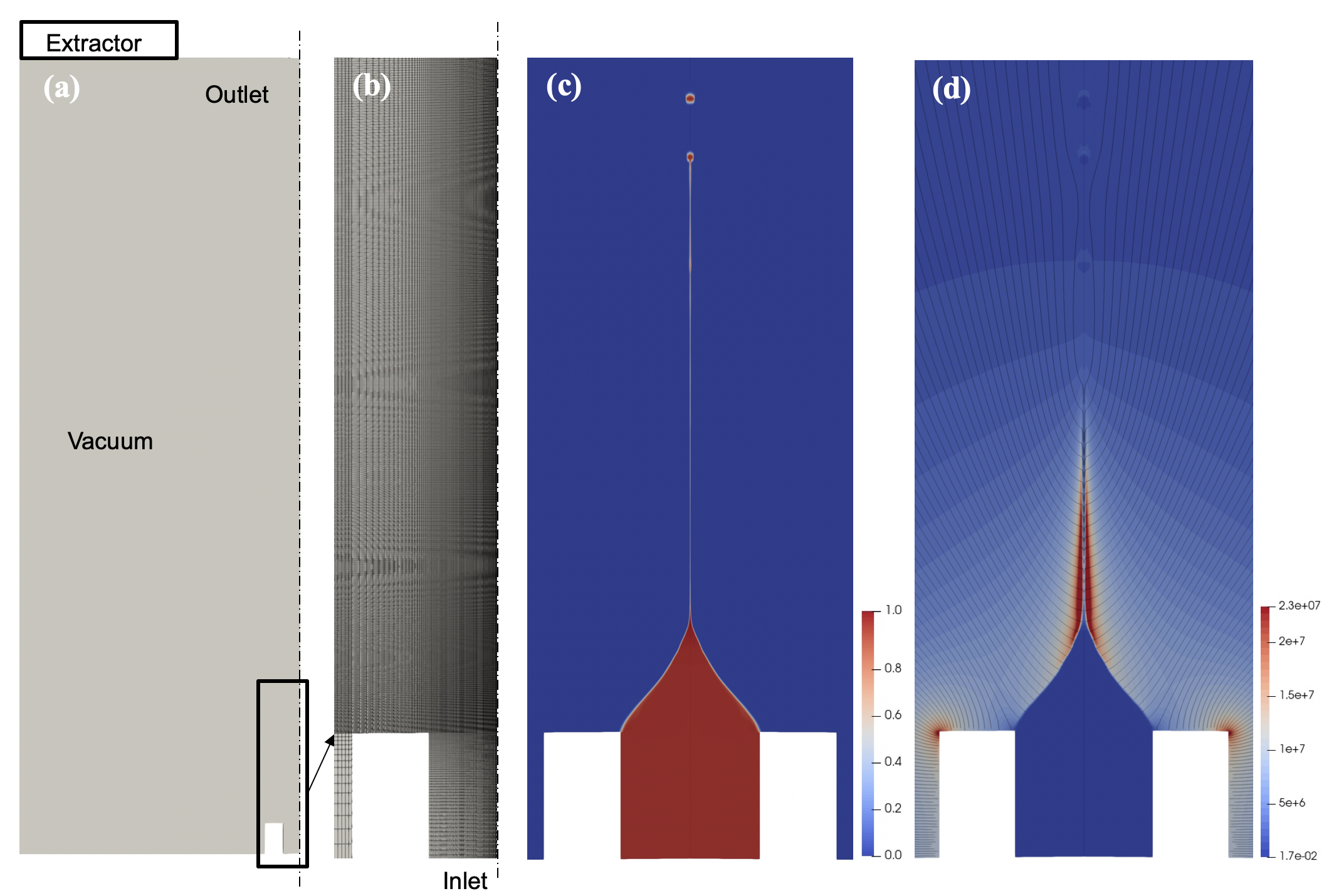}}
 \caption{Electrospray emission results for high conductivity ($\sigma=$~\SI{2.3e-4}{\siemens \per \meter}) (a) Full computational domain with 80,300 mesh cells (b) Magnified computational domain near emission region (c) Volume fraction result at steady state ($t=$~\SI{0.5}{\milli\second}) (d) Electric field magnitude.}
\label{fig:kd6}
\end{figure}

For the high conductivity case shown in Figure~\ref{fig:kd6} following the experimental setup from~\cite{gamero}, the nozzle inner and outer diameters are \SI{110}{\micro \meter} and \SI{230}{\micro \meter} respectively, with the emitter-extractor length at \SI{2.5}{\milli \meter}, and extractor orifice diameter of \SI{0.8}{\milli\meter}. Emission simulations for high conductivity liquid are more challenging than low or moderate conductivity simulations due to the near-instantaneous charge relaxation. By employing~\cref{eq:2.11,eq:2.12} to confine the charge inside the volume fraction, accurate modeling of charge transport in high conductivity liquid is possible. Due to the small charge relaxation time with high conductivity, the charge instantaneously relaxes along the meniscus ($t_e$$\ll$$t_h$) and charge transport by Ohmic conduction dominates over surface charge convection. Note that the maximum electric field at the cone-to-jet region is \SI{4.1e7}{\volt \per \meter}, well below the critical electric field required for electric field-induced ion emission (\SI{\sim e9}{\volt \per \meter}, \cite{gamero_ionemission}) indicating that ion emission is a negligible component of charge transport here. As shown in Figure~\ref{fig:kd6}, high electrical conductivity encourages the charge to accumulate at the very tip of the cone-jet. Charge accumulation at the tip allows a convex-concave meniscus to form for tributyl phosphate (TBP), unlike the concave meniscus calculated and observed for heptane (see Figure~\ref{fig:kd2}), where a relatively high emitter voltage was required to reach a steady cone-jet, when compared with that of the high conductivity liquid. With even higher conductivity liquids, such as the ionic liquid, EMI-Im (1-Ethyl-3-methylimidazolium bis(trifluoromethylsulfonyl)imide), the meniscus becomes a more convex conical shape with the Taylor cone appearing only at the cone-to-jet region, where cone-to-jet length is relatively short.


\begin{figure}
  \centerline{\includegraphics[scale=0.15]{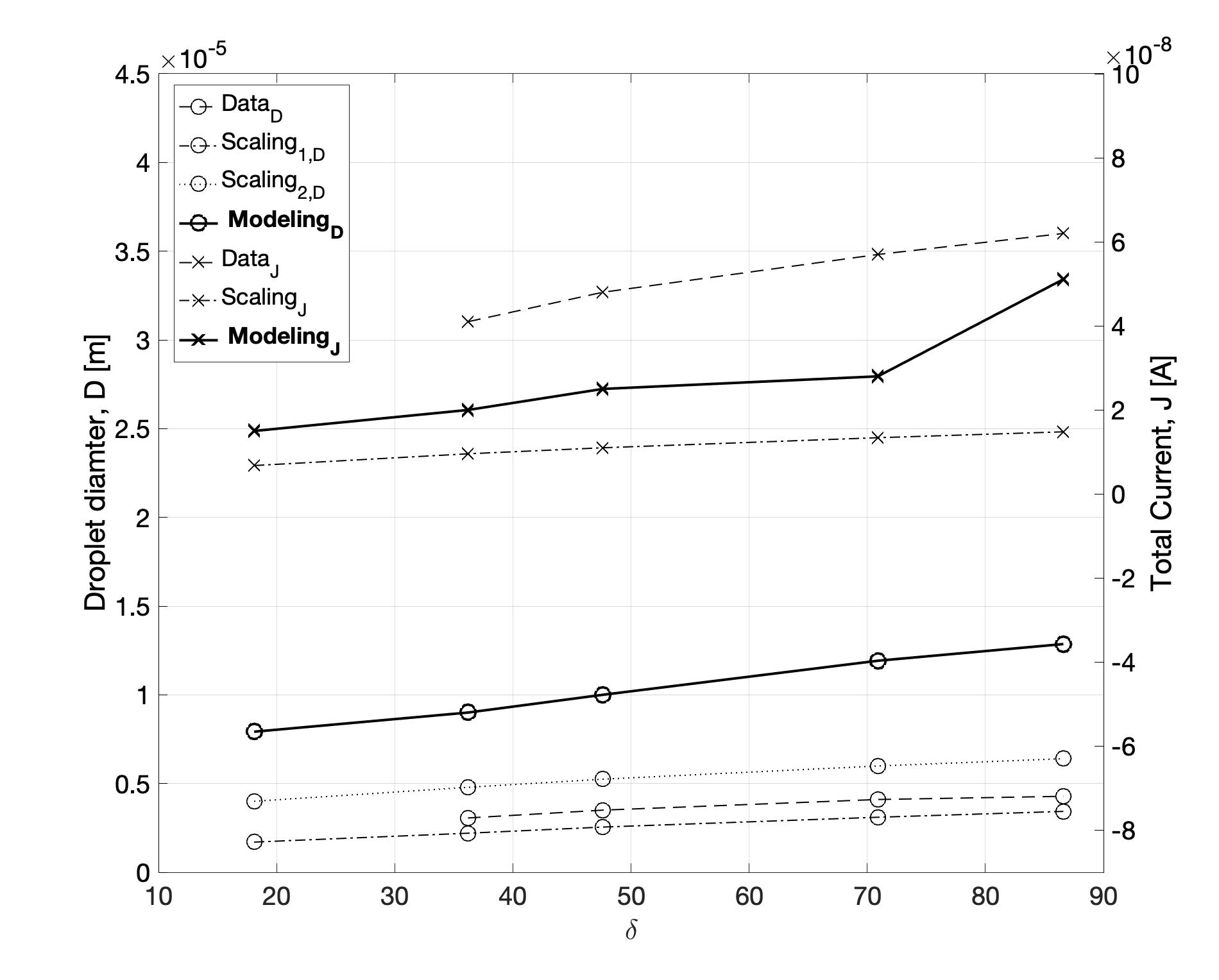}}
\caption{Quantitative validation of high electrical conductivity liquid, tributyl phosphate (TBP).}
\label{fig:kd7}
\end{figure}

Droplet diameter and total current are measured from the model to be compared with experimental values~(\cite{gamero}) and the scaling laws from~\cite{ganan_scaling},
\begin{align}\label{eq:3.3}         
d = (\rho\varepsilon_{0}Q^3/\sigma\gamma)^{1/6},\\
    I = (\gamma \sigma Q)^{1/2},
\end{align}
and
\begin{equation}\label{eq:3.5}
    d = (\varepsilon_{r}\varepsilon_{0}Q/\sigma)^{1/3}
\end{equation}
from \cite{delamora} as a function of $\delta$. In Figure~\ref{fig:kd7}, the model is well validated with good agreement between calculated droplet diameters and the total currents with the given experimental setup and operating conditions. Also shown in Figure~\ref{fig:kd7}, the model agrees with experimental trends of total current increasing with flow rate to the half power. Differences between the model and the experimental results for droplet diameter may indicate numerical uncertainty or the influence of a viscous dissipated self-heating at relatively low Reynolds number where a temperature-dependent electrical conductivity allows higher electrostatic force(\cite{tempEMIIM,Gamerodissipation}). 

\begin{figure}
  \centerline{\includegraphics[scale=0.28]{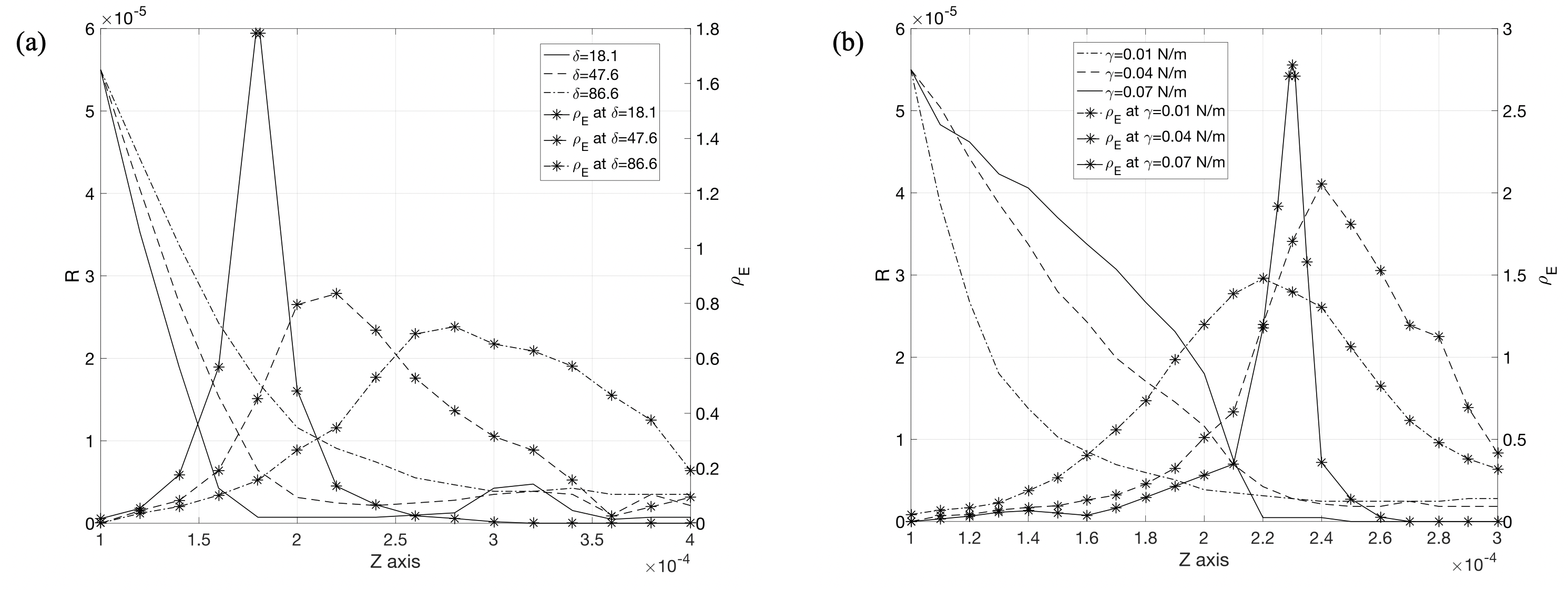}}
  \caption{Meniscus and charge distribution for different (a) flow rate and (b) surface tension.}
\label{fig:kd8}
\end{figure}

Further, for high conductivity liquid, the shape of the meniscus and the charge density along the surface is investigated by changing $\delta$ and $\gamma$ in Figure~\ref{fig:kd8}. The location of the surface, $R$, has a steeper gradient with $\delta=18.1$, which allows a smaller jet diameter and shorter jet length due to the high charge concentration. Ga\~n\'an-Calvo obtained the approximate universal expression for the surface charge, $q_s$, on the jet at the breakup point as: 

\begin{equation}\label{eq:3.6}
    q_s=\varepsilon_0 E_0 = 0.62 (\varepsilon_0 \gamma^2 \rho \sigma ^2)^{1/6},
\end{equation}

from a quasi-one-dimensional analytical model. Equation~\ref{eq:3.6} suggests that the surface charge is independent of the jet size and liquid flow rate for high enough flow rates (\cite{ganan_surfacecharge}). Results from the presented model also indicate a relatively consistent maximum charge density, $\rho_E=$~\SI{0.8}, and similar profiles for high flow rates over $\delta~=~46$, for both moderate and high conductivity liquids. However, when the flow rate reaches a minimum stable flow rate, $\delta = 18.1$, the maximum charge density greatly increases to $\rho_E$~\SI{\sim 2} with a shorter cone-to-jet length of $L_{cj}=$~\SI{65}{\micro\meter}. This indicates that in the cone-to-jet region, the dependence of charge density on jet size, and flow rate would also affect the charge-to-mass ratio of emitted droplets. An increased normal electric field ($E_n$) at the cone-to-jet region that results from increased charge concentration at the minimum stable flow rate could explain the observation of ion evaporation at high conductivity limits(\cite{ionemission,gamero_direct_measurement,garoz_ionemission,romero_ionemission}).

Similarly, a surface tension sensitivity analysis shows the transition of the meniscus from concave to convex, allowing for high charge concentration at $\gamma=\SI{0.07}{\newton \per \meter}$, shown in Figure~\ref{fig:kd8}(b). Allowing high charge concentration due to high surface tension validates the hybrid, experimental-analytical relations of charge density from \cite{ganan_surfacecharge}, where $\gamma^{\frac{1}{3}}$ is proportional to the charge density. We suggest this correlation is due to the convex meniscus where the cone-to-jet length is extremely short, which allows greater charge concentration up to $\rho_E=$~\SI{2.8} for $\gamma=\SI{0.07}{\newton \per \meter}$. The convex meniscus due to higher surface tension ($\gamma >$~\SI{0.05}{\newton \per \meter}) followed by the shortening of the cone-to-jet length allows for higher charge accumulation. The convex meniscus is only achievable if the high-conductivity-induced electric force surpasses the surface  tension and viscous forces of the liquid. Applying high voltage, on the other hand, results in a high electrostatic force acting at the vicinity of the emitter tip. High electrostatic force at the tip causes a concave meniscus with larger cone-to-jet length, and subsequently decentralizes the charge accumulation, thus making it difficult to obtain high droplet specific charge, even with a high electrostatic force.

\begin{figure}
  \centerline{\includegraphics[scale=0.23]{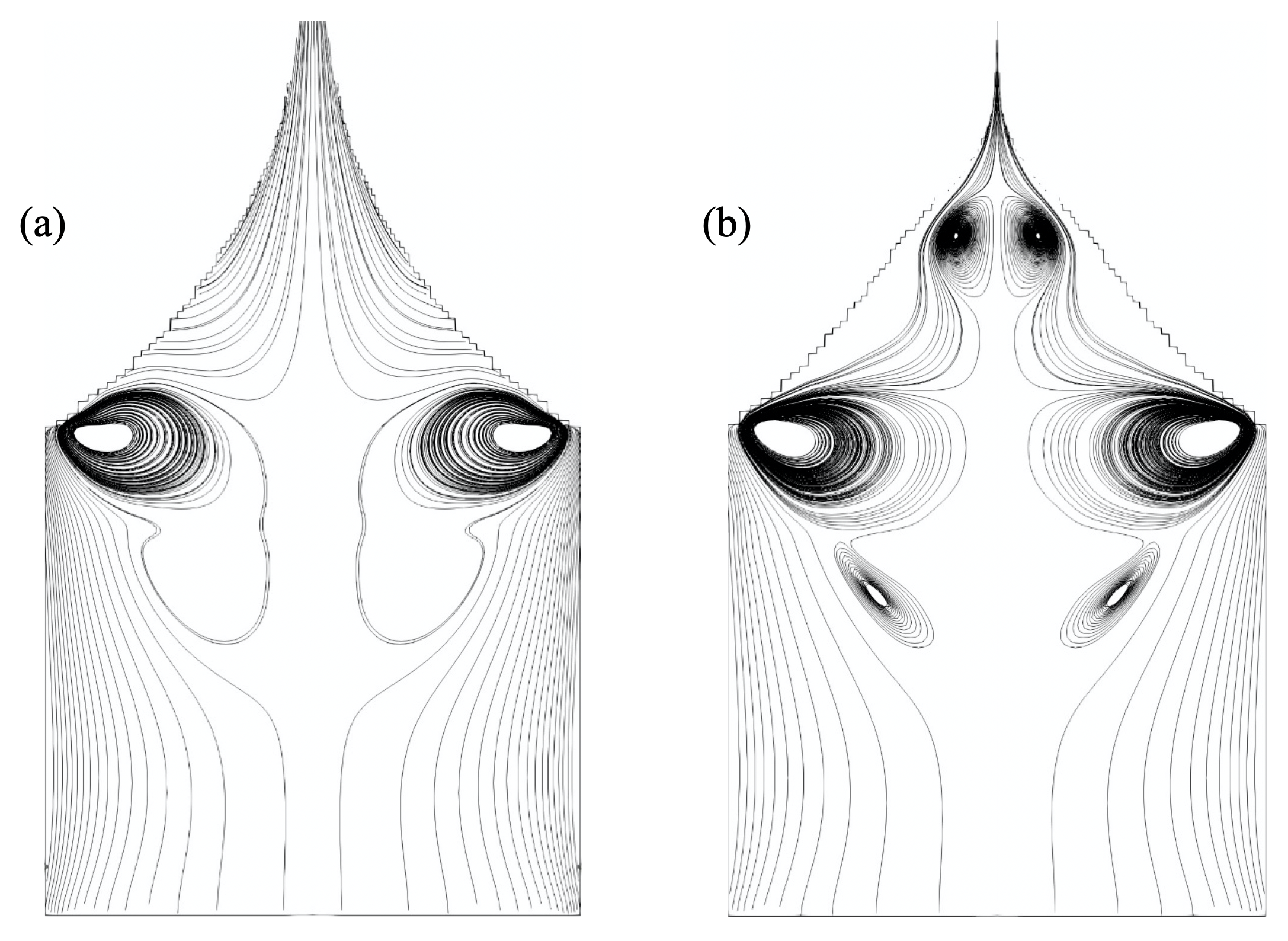}}
  \caption{Recirculation flow at $\delta~=~18.1, Re_E~=~0.86$ (a) Steady cone-jet mode ($t=$~\SI{0.1}{\milli\second}) (b) Onset of pulsating mode ($t=$~\SI{0.5}{\milli\second}) near minimum flow rate where whipping motion occurs at the cone-to-jet region with the pulsation of the jet in the experiment.}
\label{fig:kd9}
\end{figure}

An interesting finding for high conductivity liquid emission is the presence and structure of recirculation flows. In Figure~\ref{fig:kd9}, the transient emission begins with (a) a nominal cone-jet with one dominant recirculation cell at $t=$~\SI{0.1}{\milli\second}; and ends with (b) an  additional strong recirculation at the cone tip at $t=$~\SI{0.5}{\milli\second}. Unlike for moderate conductivity, two or more recirculation cells emerge inside the bulk high conductivity liquid near the minimum stable flow rate, for $\delta = 18.1, Re_E=0.86$ at the end of each pulsation cycle in Figure~\ref{fig:kd9}. Growth of recirculations indicates that the high Coulombic force induced from the small $\delta$ could be attributed to the onset of axisymmetric instabilities, resulting in a spontaneous toroidal motion of the flow as the axisymmetric instabilities breaks down (\cite{shtern,shtern_striking,barrero_swirl}). For example, Axisymmetric meridional motion of the flow intensifies in the advent of more than one recirculation cell with lower $\delta$ and higher $Re_E$ in the high conductivity regime. This implies that the onset of axisymmetric instability due to the high tangential electrostatic force at the cone-to-jet region results in a toroidal motion and intensifies along the jet. The strength of this toroidal motion increases with the applied potential (\cite{gupta}).

\begin{figure}
  \centerline{\includegraphics[scale=0.25]{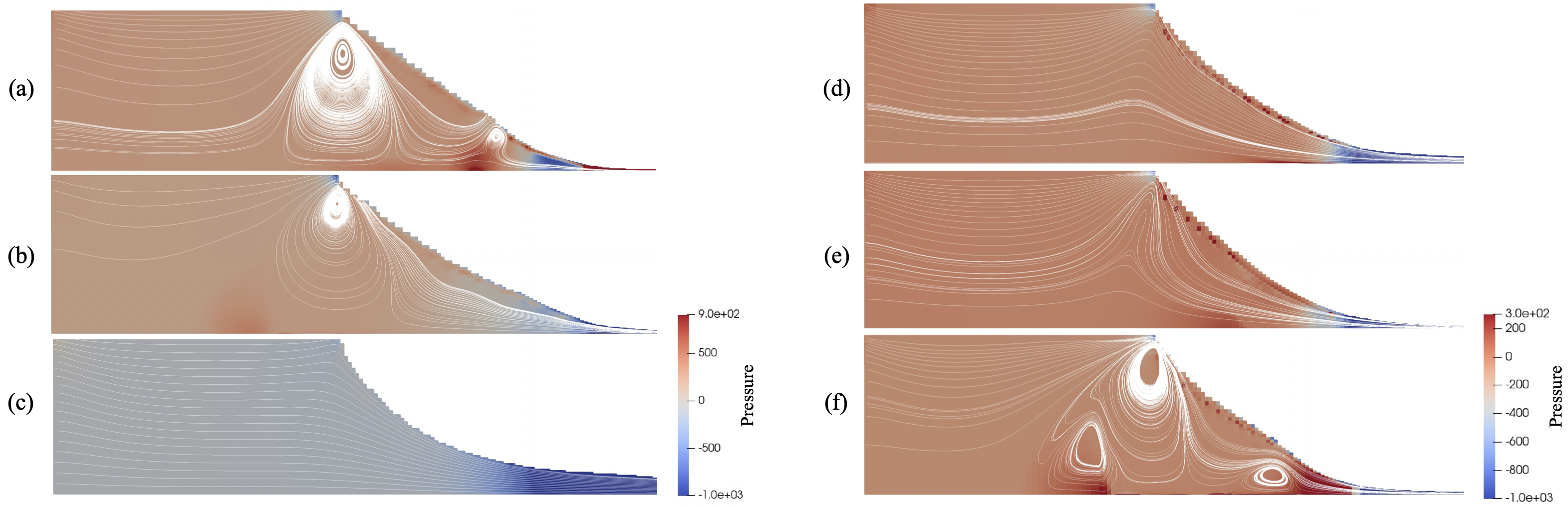}}
  \caption{Streamlined recirculation flow for different $\delta, Re_E$ (a) $\delta=18.1, (b) \delta=47.6, (c) \delta=86.6, (d) Re_E=0.22, (e) Re_E=0.43, (f) Re_E=1.72$.}
\label{fig:kd10}
\end{figure}

Recirculation for various $\delta$ and $Re_E$ is investigated in Figure~\ref{fig:kd10}. The recirculation cell strengthens with decreasing flow rate and additional steady recirculations present near the cone-tip at $\delta=18.1$. Increasing $Re_E$ also allows the formation of additional recirculation cells. Increasing the viscosity of the liquid stabilizes the flow emission and flattens the jet profile. This verifies the surface tension effect on tip-streaming instability where surfactants weaken the surface tension and subside the instability of converging flow (\cite{tseng}).



   \begin{figure}
  \centerline{\includegraphics[scale=0.25]{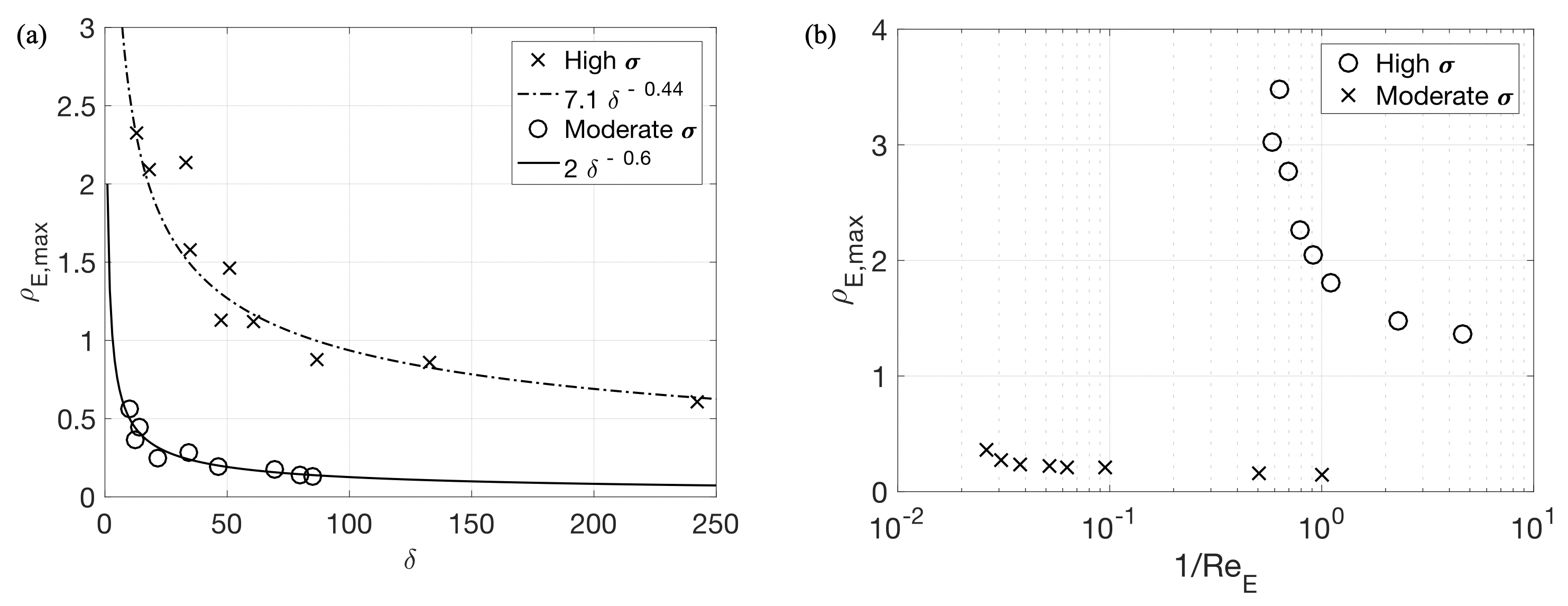}}
  \caption{Maximum charge density at cone-to-jet region (a) as a function of the non-dimensional flow rate $\delta$ and (b) as a function of the electrical Reynolds number $1/Re_E$.}
\label{fig:kd11}
\end{figure}

\begin{figure}
  \centerline{\includegraphics[scale=0.28]{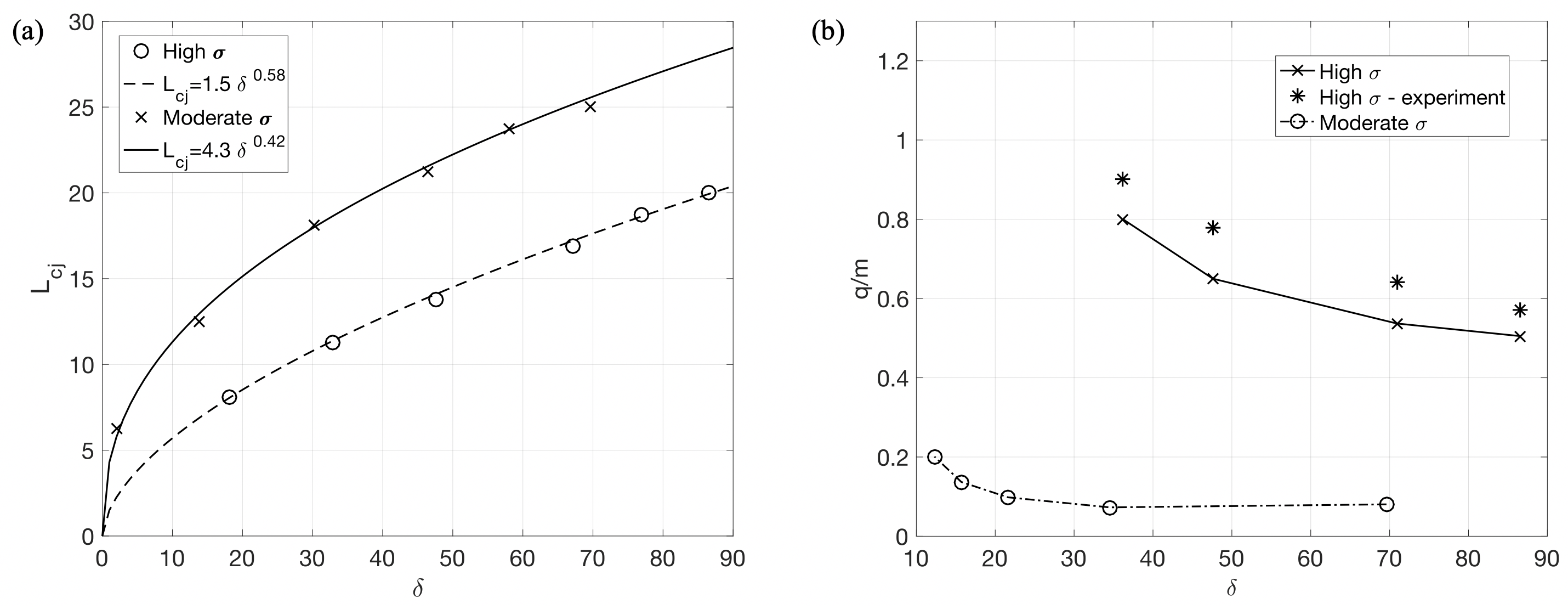}}
\caption{(a) Cone-to-jet length and (b) Charge-to-mass ratio of the emitted droplet as a function of the non-dimensional flow rate $\delta$}
\label{fig:kd12}
\end{figure}

Charge distribution along the meniscus has been of interest for a number of years (\cite{gamero2019,higuera,ganan_surfacecharge}. As shown in Figure~\ref{fig:kd11}, we explored the maximum charge densities across various non-dimensional flow rates as calculated by the model. A power-law curve fit yields $\rho_{E,max} \sim 2\delta^{-0.6}$ for moderate conductivity liquid, heptane. Comparing to Figure~\ref{fig:kd12}(a), we see that for moderate conductivity, this weaker dependence of delta weakens the charge concentration at the cone-to-jet region, thus relatively lengthening $L_{cj}$ to the proportion of $L_{cj} \sim \delta^{0.42}$. In contrast, the high conductivity case with tributyl phosphate ($Re_E=0.86$) yields $\rho_{E,max} \sim 7.1\delta^{-0.44}$ shortening $L_{cj}$ to the proportion of $L_{cj} \sim \delta^{0.58}$. Furthermore, the results show stronger dependence than $L_{cj} \sim\delta^{0.17}$ from \cite{gamero2019} suggesting that the convex meniscus allows higher charge concentration and shorter $L_{cj}$. Similarly, the maximum charge density on varying reciprocal $Re_E=(\frac{\rho \varepsilon_0 \gamma^2}{\mu^3 \sigma})^{1/3}$ is shown in Figure~\ref{fig:kd11}(b). Increasing $Re_E$ (lower 1/$Re_E$) allows higher charge concentration and this effect intensifies with higher conductivity. 
Ultimately, the charge-to-mass ratio of the emitted droplets decreases with increasing $\delta$ in Figure~\ref{fig:kd12}(b) where the model validates the experimental observations for high conductivity liquid.

\section{Conclusion}\label{sec4}
We have developed the electrohydrodynamic equations with finite volume analysis specifically for moderate to high conductivity electrosprya liquids. The model has been validated against experiments and revealed of confirmed important emission behavior for these liquid regimes. When considering instantaneous charge relaxation of high conductivity liquids, the surface charge convection effect is negligible, allowing conservation of the charge of the emitted droplets. Measurements of charge-to-mass ratio, droplet diameters, and total current from the model show good agreement with experimental observations for both moderate and high conductivity fluids. One important observation that can be made from the model results is that the charge distribution along the meniscus is exclusively determined by the cone shape and vice versa, such that lower $\delta$ and higher $Re_E$ provide higher maximum charge density. In particular, higher conductivity ($\sigma$) allows exceeding the minimum Coulombic force that impedes cone-jet generation due to the higher surface tension ($\gamma$) where higher $\gamma$ gives higher charge density. High charge concentration at the meniscus leads to a higher charge-to-mass ratio of emitted droplets with higher jetting velocity. This describes the convex meniscus typically observed for higher conductivity electrospray emission, while a concave meniscus are more indicative of liquids in the moderate to low conductivity regime.

While the recirculation flow inside a moderately-conductive bulk liquid has been well characterized in previous studies, our simulations show that higher conductivity liquid yield additional recirculation cells near the cone-to-jet region at lower $\delta$ and higher $Re_E$. Additional axisymmetric recirculations indicate the onset of a flow instability caused by the high electrostatic force resulting from high charge concentration at the cone-to-jet region, leading to a breakdown of the axisymmetric recirculation at the end of the pulsating cycle. The shape of the meniscus and the concavity of the cone determines the length of the cone-to-jet region and the charge distribution along the meniscus and vice versa. The length decreases with $\delta$ and presents stronger dependence on conductivity leading to the higher charge of the emitted droplet.

\section{Acknowledgements}\label{sec5}

This effort was supported by a grant from NASA's Jet Propulsion Laboratory, California Institute of Technology to support the LISA CMT development plan(NASA/JPL Award No. 1580267), the Air Force Research Lab at Edwards AFB, CA(Award No. 16-EPA-RQ-09), and Air Force Office of Scientific Research(AFOSR)(Award No. FA9550-21-1-0067). The authors would like to thank John Ziemer from NASA JPL; and David Bilyeu and Daniel Eckhardt from Air force Research Laboratory; and Adam Collins from UCLA for their insightful discussions.

\section{Declaration of Interests}\label{sec6}
The authors report no conflict of interest.

\bibliographystyle{jfm}
\bibliography{jfm-instructions}

\begin{thebibliography}{43}
\expandafter\ifx\csname natexlab\endcsname\relax\def\natexlab#1{#1}\fi
\def\au#1{#1} \def\ed#1{#1} \def\yr#1{#1}\def\at#1{#1}\def\jt#1{\textit{#1}}
  \def\bt#1{#1}\def\bvol#1{\textbf{#1}} \def\vol#1{#1} \def\pg#1{#1}
  \def\publ#1{#1}\def\arxiv#1{#1}\def\org#1{#1}\def\st#1{\textit{#1}}

\bibitem[Barrero {\em et~al.\/}(1998)Barrero, Ga\~n\'an Calvo, D\'avila,
  Palacio \& G\'omez-Gonz\'alez]{barrero_swirl}
{\sc \au{Barrero, A.}, \au{Ga\~n\'an Calvo, A.~M.}, \au{D\'avila, J.},
  \au{Palacio, A.} \& \au{G\'omez-Gonz\'alez, E.}} \yr{1998}  \at{Low and high
  reynolds number flows inside taylor cones}.  \jt{Phys. Rev. E}  \bvol{58},
  \pg{7309--7314}.

\bibitem[Brackbill {\em et~al.\/}(1992)Brackbill, Kothe \&
  Zemach]{BRACKBILL1992335}
{\sc \au{Brackbill, J.U}, \au{Kothe, D.B} \& \au{Zemach, C}} \yr{1992}  \at{A
  continuum method for modeling surface tension}.  \jt{Journal of Computational
  Physics}  \bvol{100}~(2),  \pg{335--354}.

\bibitem[Ga\~n\'an Calvo(1999)]{ganan_surfacecharge}
{\sc \au{Ga\~n\'an Calvo, Alfonso~M}} \yr{1999}  \at{The surface charge in
  electrospraying: Its nature and its universal scaling laws}.  \jt{Journal of
  Aerosol Science}  \bvol{30}~(7),  \pg{863--872}.

\bibitem[Ga\~n\'an Calvo(2004)]{ganan_scaling}
{\sc \au{Ga\~n\'an Calvo, Alfonso~M.}} \yr{2004}  \at{On the general scaling
  theory for electrospraying}.  \jt{Journal of Fluid Mechanics}  \bvol{507},
  \pg{203–212a}.

\bibitem[Ga\~n\'an Calvo \& Montanero(2009{\natexlab{{\em
  a\/}}})]{ganan_montanero2009}
{\sc \au{Ga\~n\'an Calvo, Alfonso~M.} \& \au{Montanero, Jos\'e~M.}}
  \yr{2009{\natexlab{{\em a\/}}}}  \at{Revision of capillary cone-jet physics:
  Electrospray and flow focusing}.  \jt{Phys. Rev. E}  \bvol{79},  \pg{066305}.

\bibitem[Ga\~n\'an Calvo \& Montanero(2009{\natexlab{{\em
  b\/}}})]{ganan_recirculation}
{\sc \au{Ga\~n\'an Calvo, Alfonso~M.} \& \au{Montanero, Jos\'e~M.}}
  \yr{2009{\natexlab{{\em b\/}}}}  \at{Revision of capillary cone-jet physics:
  Electrospray and flow focusing}.  \jt{Phys. Rev. E}  \bvol{79},  \pg{066305}.

\bibitem[Cherney(1999)]{cherney_1999_structure}
{\sc \au{Cherney, Leonid~T.}} \yr{1999}  \at{Structure of taylor cone-jets:
  limit of low flow rates}.  \jt{Journal of Fluid Mechanics}  \bvol{378},
  \pg{167–196}.

\bibitem[Dastourani {\em et~al.\/}(2018)Dastourani, Jahannama \&
  Eslami-Majd]{dastourani}
{\sc \au{Dastourani, H.}, \au{Jahannama, M.R.} \& \au{Eslami-Majd, A.}}
  \yr{2018}  \at{A physical insight into electrospray process in cone-jet mode:
  Role of operating parameters}.  \jt{International Journal of Heat and Fluid
  Flow}  \bvol{70},  \pg{315--335}.

\bibitem[De~La~Mora \& Loscertales(1994)]{delamora}
{\sc \au{De~La~Mora, J.~Fernández} \& \au{Loscertales, I.~G.}} \yr{1994}
  \at{The current emitted by highly conducting taylor cones}.  \jt{Journal of
  Fluid Mechanics}  \bvol{260},  \pg{155–184}.

\bibitem[Feng(1999)]{feng}
{\sc \au{Feng, James~Q.}} \yr{1999}  \at{Electrohydrodynamic behaviour of a
  drop subjected to a steady uniform electric field at finite electric reynolds
  number}.  \jt{Proceedings: Mathematical, Physical and Engineering Sciences}
  \bvol{455}~(1986),  \pg{2245--2269}.

\bibitem[Gamero-Casta\~no(2002)]{gamero_ionemission}
{\sc \au{Gamero-Casta\~no, Manuel}} \yr{2002}  \at{Electric-field-induced ion
  evaporation from dielectric liquid}.  \jt{Phys. Rev. Lett.}  \bvol{89},
  \pg{147602}.

\bibitem[Gamero-Casta\~no \& Hruby(2002)]{gamero}
{\sc \au{Gamero-Casta\~no, Manuel} \& \au{Hruby, Vladimir}} \yr{2002}
  \at{Electric measurements of charged sprays emitted by cone-jets}.
  \jt{Journal of Fluid Mechanics}  \bvol{459},  \pg{245–276}.

\bibitem[Gamero-Casta{\~n}o(2019)]{Gamerodissipation}
{\sc \au{Gamero-Casta{\~n}o, Manuel}} \yr{2019}  \at{Dissipation in cone-jet
  electrosprays and departure from isothermal operation.}  \jt{Physical review.
  E}  \bvol{99 6-1},  \pg{061101}.

\bibitem[Gamero-Castaño \& Magnani(2019)]{gamero2019}
{\sc \au{Gamero-Castaño, M.} \& \au{Magnani, M.}} \yr{2019}  \at{Numerical
  simulation of electrospraying in the cone-jet mode}.  \jt{Journal of Fluid
  Mechanics}  \bvol{859},  \pg{247–267}.

\bibitem[Gamero-Castaño \& Fernández de~la
  Mora(2000)]{gamero_direct_measurement}
{\sc \au{Gamero-Castaño, M.} \& \au{Fernández de~la Mora, J.}} \yr{2000}
  \at{Direct measurement of ion evaporation kinetics from electrified liquid
  surfaces}.  \jt{The Journal of Chemical Physics}  \bvol{113}~(2),
  \pg{815--832},  \arxiv{arXiv: https://doi.org/10.1063/1.481857}.

\bibitem[Ga{\~{n}}{\'{a}}n-Calvo {\em et~al.\/}(2013)Ga{\~{n}}{\'{a}}n-Calvo,
  Rebollo-Mu{\~{n}}oz \& Montanero]{ganan_polar}
{\sc \au{Ga{\~{n}}{\'{a}}n-Calvo, A~M}, \au{Rebollo-Mu{\~{n}}oz, N} \&
  \au{Montanero, J~M}} \yr{2013}  \at{The minimum or natural rate of flow and
  droplet size ejected by taylor cone{\textendash}jets: physical symmetries and
  scaling laws}.  \jt{New Journal of Physics}  \bvol{15}~(3),  \pg{033035}.

\bibitem[Garoz {\em et~al.\/}(2007)Garoz, Bueno, Larriba, Castro, Romero-Sanz,
  Fernandez de~la Mora, Yoshida \& Saito]{garoz_ionemission}
{\sc \au{Garoz, D.}, \au{Bueno, C.}, \au{Larriba, C.}, \au{Castro, S.},
  \au{Romero-Sanz, I.}, \au{Fernandez de~la Mora, J.}, \au{Yoshida, Y.} \&
  \au{Saito, G.}} \yr{2007}  \at{Taylor cones of ionic liquids from capillary
  tubes as sources of pure ions: The role of surface tension and electrical
  conductivity}.  \jt{Journal of Applied Physics}  \bvol{102}~(6),
  \pg{064913},  \arxiv{arXiv: https://doi.org/10.1063/1.2783769}.

\bibitem[Gañán-Calvo {\em et~al.\/}(1997)Gañán-Calvo, Dávila \&
  Barrero]{GANANCALVOscaling1997}
{\sc \au{Gañán-Calvo, A.M.}, \au{Dávila, J.} \& \au{Barrero, A.}} \yr{1997}
  \at{Current and droplet size in the electrospraying of liquids. scaling
  laws}.  \jt{Journal of Aerosol Science}  \bvol{28}~(2),  \pg{249--275}.

\bibitem[Gañán-Calvo {\em et~al.\/}(2011)Gañán-Calvo, Ferrera, Torregrosa,
  Herrada \& Marchand]{ganan_rec}
{\sc \au{Gañán-Calvo, A.~M.}, \au{Ferrera, C.}, \au{Torregrosa, M.},
  \au{Herrada, M.~A.} \& \au{Marchand, M.}} \yr{2011}  \at{Experimental and
  numerical study of the recirculation flow inside a liquid meniscus focused by
  air}.  \jt{Microfluidics and Nanofluidics}  \bvol{11}~(1),  \pg{65--74}.

\bibitem[Gupta {\em et~al.\/}(2019)Gupta, Mishra \& Panigrahi]{gupta}
{\sc \au{Gupta, A.}, \au{Mishra, B.~K.} \& \au{Panigrahi, P.}} \yr{2019}
  \bt{Experimental investigation of flow field inside a taylor cone}.

\bibitem[Hayati {\em et~al.\/}(1986)Hayati, Bailey \& Tadros]{hayati}
{\sc \au{Hayati, I.}, \au{Bailey, A.~I.} \& \au{Tadros, Th.~F.}} \yr{1986}
  \at{Mechanism of stable jet formation in electrohydrodynamic atomization}.
  \jt{Nature}  \bvol{319}~(6048),  \pg{41--43}.

\bibitem[Herrada {\em et~al.\/}(2012)Herrada, L\'opez-Herrera, Ga\~n\'an Calvo,
  Vega, Montanero \& Popinet]{herrada}
{\sc \au{Herrada, M.~A.}, \au{L\'opez-Herrera, J.~M.}, \au{Ga\~n\'an Calvo,
  A.~M.}, \au{Vega, E.~J.}, \au{Montanero, J.~M.} \& \au{Popinet, S.}}
  \yr{2012}  \at{Numerical simulation of electrospray in the cone-jet mode}.
  \jt{Phys. Rev. E}  \bvol{86},  \pg{026305}.

\bibitem[Higuera(2003)]{higuera}
{\sc \au{Higuera, F.~J.}} \yr{2003}  \at{Flow rate and electric current emitted
  by a taylor cone}.  \jt{Journal of Fluid Mechanics}  \bvol{484},
  \pg{303–327}.

\bibitem[Lanauze {\em et~al.\/}(2015)Lanauze, Walker \& Khair]{lanauze}
{\sc \au{Lanauze, Javier~A.}, \au{Walker, Lynn~M.} \& \au{Khair, Aditya~S.}}
  \yr{2015}  \at{Nonlinear electrohydrodynamics of slightly deformed oblate
  drops}.  \jt{Journal of Fluid Mechanics}  \bvol{774},  \pg{245–266}.

\bibitem[L{\'o}pez-Herrera {\em et~al.\/}(2011)L{\'o}pez-Herrera, Popinet \&
  Herrada]{lopez}
{\sc \au{L{\'o}pez-Herrera, J.}, \au{Popinet, S.} \& \au{Herrada, M.}}
  \yr{2011}  \at{A charge-conservative approach for simulating
  electrohydrodynamic two-phase flows using volume-of-fluid}.  \jt{J. Comput.
  Phys.}  \bvol{230},  \pg{1939--1955}.

\bibitem[Melcher(1981)]{melcher_1981}
{\sc \au{Melcher, James~R.}} \yr{1981} {\em Continuum electromechanics\/}.
  \publ{MIT Press}.

\bibitem[Melcher \& Taylor(1969)]{melcher}
{\sc \au{Melcher, J~R} \& \au{Taylor, G~I}} \yr{1969}
  \at{Electrohydrodynamics: A review of the role of interfacial shear
  stresses}.  \jt{Annual Review of Fluid Mechanics}  \bvol{1}~(1),
  \pg{111--146},  \arxiv{arXiv:
  https://doi.org/10.1146/annurev.fl.01.010169.000551}.

\bibitem[Miller {\em et~al.\/}(2014)Miller, Prince, Bemish \&
  Rovey]{ionemission}
{\sc \au{Miller, Shawn~W.}, \au{Prince, Benjamin~D.}, \au{Bemish, Raymond~J.}
  \& \au{Rovey, Joshua~L.}} \yr{2014}  \at{Electrospray of
  1-butyl-3-methylimidazolium dicyanamide under variable flow rate operations}.
   \jt{Journal of Propulsion and Power}  \bvol{30}~(6),  \pg{1701--1710},
  \arxiv{arXiv: https://doi.org/10.2514/1.B35170}.

\bibitem[Popinet(2003)]{popinet}
{\sc \au{Popinet, Stéphane}} \yr{2003}  \at{Gerris: a tree-based adaptive
  solver for the incompressible euler equations in complex geometries}.
  \jt{Journal of Computational Physics}  \bvol{190}~(2),  \pg{572--600}.

\bibitem[Roghair {\em et~al.\/}(2015)Roghair, Musterd, van~den Ende, Kleijn,
  Kreutzer \& Mugele]{ivo}
{\sc \au{Roghair, Ivo}, \au{Musterd, Michiel}, \au{van~den Ende, Dirk},
  \au{Kleijn, Chris}, \au{Kreutzer, Michiel} \& \au{Mugele, Frieder}} \yr{2015}
   \at{A numerical technique to simulate display pixels based on
  electrowetting}.  \jt{Microfluidics and Nanofluidics}  \bvol{19}~(2),
  \pg{465--482}.

\bibitem[Romero-Sanz {\em et~al.\/}(2003)Romero-Sanz, Bocanegra, Fernandez
  de~la Mora \& Gamero-Castaño]{romero_ionemission}
{\sc \au{Romero-Sanz, I.}, \au{Bocanegra, R.}, \au{Fernandez de~la Mora, J.} \&
  \au{Gamero-Castaño, M.}} \yr{2003}  \at{Source of heavy molecular ions based
  on taylor cones of ionic liquids operating in the pure ion evaporation
  regime}.  \jt{Journal of Applied Physics}  \bvol{94}~(5),  \pg{3599--3605},
  \arxiv{arXiv: https://doi.org/10.1063/1.1598281}.

\bibitem[Salipante \& Vlahovska(2010)]{salipante}
{\sc \au{Salipante, {Paul F.}} \& \au{Vlahovska, {Petia M.}}} \yr{2010}
  \at{Electrohydrodynamics of drops in strong uniform dc electric fields}.
  \jt{Physics of Fluids}  \bvol{22}~(11), funding Information: Acknowledgment
  is made to the Donors of the American Chemical Society Petroleum Research
  Fund for partial support of this research. Partial funding has also been
  provided by NSF CAREER award CBET–0846247. We thank the referees for useful
  comments.

\bibitem[Saville(1997)]{saville}
{\sc \au{Saville, D.~A.}} \yr{1997}  \at{Electrohydrodynamics: The
  taylor-melcher leaky dielectric model}.  \jt{Annual Review of Fluid
  Mechanics}  \bvol{29}~(1),  \pg{27--64},  \arxiv{arXiv:
  https://doi.org/10.1146/annurev.fluid.29.1.27}.

\bibitem[Sengupta {\em et~al.\/}(2017)Sengupta, Walker \& Khair]{sengupta}
{\sc \au{Sengupta, Rajarshi}, \au{Walker, Lynn~M.} \& \au{Khair, Aditya~S.}}
  \yr{2017}  \at{The role of surface charge convection in the
  electrohydrodynamics and breakup of prolate drops}.  \jt{Journal of Fluid
  Mechanics}  \bvol{833},  \pg{29–53}.

\bibitem[Shtern(2012)]{shtern}
{\sc \au{Shtern, Vladimir}} \yr{2012} {\em Bifurcation of Swirl in Conical
  Counterflows\/},  \pg{p. 28–59}.  \publ{Cambridge University Press}.

\bibitem[Shtern \& Barrero(1994)]{shtern_striking}
{\sc \au{Shtern, Vladimir} \& \au{Barrero, Antonio}} \yr{1994}  \at{Striking
  features of fluid flows in taylor cones related to electrosprays}.
  \jt{Journal of Aerosol Science}  \bvol{25}~(6),  \pg{1049--1063}.

\bibitem[Tang \& Gomez(1996)]{tang}
{\sc \au{Tang, Keqi} \& \au{Gomez, Alessandro}} \yr{1996}  \at{Monodisperse
  electrosprays of low electric conductivity liquids in the cone-jet mode}.
  \jt{Journal of Colloid and Interface Science}  \bvol{184}~(2),
  \pg{500--511}.

\bibitem[Taylor(1964)]{taylor}
{\sc \au{Taylor, Geoffrey~Ingram}} \yr{1964}  \at{Disintegration of water drops
  in an electric field}.  \jt{Proceedings of the Royal Society of London.
  Series A. Mathematical and Physical Sciences}  \bvol{280}~(1382),
  \pg{383--397},  \arxiv{arXiv:
  https://royalsocietypublishing.org/doi/pdf/10.1098/rspa.1964.0151}.

\bibitem[Tomar {\em et~al.\/}(2007)Tomar, Gerlach, Biswas, Alleborn, Sharma,
  Durst, Welch \& Delgado]{TOMAR20071267}
{\sc \au{Tomar, G.}, \au{Gerlach, D.}, \au{Biswas, G.}, \au{Alleborn, N.},
  \au{Sharma, A.}, \au{Durst, F.}, \au{Welch, S.W.J.} \& \au{Delgado, A.}}
  \yr{2007}  \at{Two-phase electrohydrodynamic simulations using a
  volume-of-fluid approach}.  \jt{Journal of Computational Physics}
  \bvol{227}~(2),  \pg{1267--1285}.

\bibitem[Tseng \& Prosperetti(2015)]{tseng}
{\sc \au{Tseng, Yu-Hau} \& \au{Prosperetti, Andrea}} \yr{2015}  \at{Local
  interfacial stability near a zero vorticity point}.  \jt{Journal of Fluid
  Mechanics}  \bvol{776},  \pg{5–36}.

\bibitem[Vila {\em et~al.\/}(2006)Vila, Ginés, Pico, Franjo, Jiménez, Varela
  \& Cabeza]{tempEMIIM}
{\sc \au{Vila, J.}, \au{Ginés, P.}, \au{Pico, J.M.}, \au{Franjo, C.},
  \au{Jiménez, E.}, \au{Varela, L.M.} \& \au{Cabeza, O.}} \yr{2006}
  \at{Temperature dependence of the electrical conductivity in emim-based ionic
  liquids: Evidence of vogel–tamman–fulcher behavior}.  \jt{Fluid Phase
  Equilibria}  \bvol{242}~(2),  \pg{141--146}.

\bibitem[Zeleny(1914)]{Zeleny_first}
{\sc \au{Zeleny, John}} \yr{1914}  \at{The electrical discharge from liquid
  points, and a hydrostatic method of measuring the electric intensity at their
  surfaces}.  \jt{Phys. Rev.}  \bvol{3},  \pg{69--91}.

\bibitem[Zeleny(1917)]{Zeleny_instability}
{\sc \au{Zeleny, John}} \yr{1917}  \at{Instability of electrified liquid
  surfaces}.  \jt{Phys. Rev.}  \bvol{10},  \pg{1--6}.

\end{thebibliography}

\end{document}